\documentclass[11pt, a4paper]{article}
\pdfoutput=1

\usepackage{jcappub} 

\usepackage{amsmath}
\usepackage{amsfonts}
\usepackage{amssymb}
\usepackage{graphicx}
\usepackage{mathrsfs}
\usepackage{latexsym}
\usepackage{color}
\usepackage{slashed, cancel}
\usepackage{hyperref}
\usepackage{bbold}

\usepackage[font=footnotesize,labelfont=bf]{caption} 

\numberwithin{equation}{section}

\definecolor{rossos}{rgb}{0.8,0.2,0.3}
\definecolor{bluscuro}{rgb}{0.15, 0.2, .85}
\definecolor{bluchiaro}{cmyk}{1,.3,0.,0.1}

\definecolor{orange}{rgb}{1,0.5,0}
\definecolor{blue}{rgb}{0,0,1}

\hypersetup{colorlinks, citecolor=bluscuro, linkcolor=bluscuro, urlcolor=bluscuro}

\makeatother   

\def\mx{m_{\rm DM}}
\def\gx{g_{\rm DM}}
\def\sigv{\langle \sigma v \rangle}
\def\sigvmin{\langle \sigma v \rangle_{*}}

 \def\be   {\begin{equation}}   \def\ee   {\end{equation}}
 \def\ba   {\begin{array}}      \def\ea   {\end{array}}
 \def\bea  {\begin{eqnarray}}   \def\eea  {\end{eqnarray}}
 \def\bean {\begin{eqnarray*}}  \def\eean {\end{eqnarray*}}
 \def\nn{\nonumber}

\title{Making the Most of the Relic Density for Dark Matter Searches at the LHC 14 TeV Run}

\author[a]{Giorgio Busoni,}
\author[a]{Andrea De Simone,}
\author[b]{Thomas Jacques,}
\author[b]{Enrico Morgante,}
\author[b]{Antonio Riotto}

\affiliation[a\,]{SISSA and INFN, Sezione di Trieste, via Bonomea 265, I-34136 Trieste, Italy}
\affiliation[b\,]{Section de Physique, Universit\'e de Gen\`eve,
24 quai E. Ansermet, CH-1211 Geneva, Switzerland}

\emailAdd{giorgio.busoni@sissa.it}
\emailAdd{andrea.desimone@sissa.it}
\emailAdd{thomas.jacques@unige.ch}
\emailAdd{enrico.morgante@unige.ch}
\emailAdd{antonio.riotto@unige.ch}

\abstract{
As the LHC continues to search for new weakly interacting particles, it is important to remember that the search is strongly motivated by the existence of dark matter. 
In view of a possible positive signal, it is essential to ask whether the newly discovered weakly interacting particle
can be be assigned the label  ``dark matter". Within a given set of simplified models and modest working assumptions, we reinterpret the relic abundance bound as a relic abundance range, and compare the parameter space yielding the correct relic abundance with projections of the Run II exclusion regions. Assuming that  dark matter is within the reach of the LHC, we also make the comparison with the potential 5$\sigma$  discovery regions. Reversing the logic, relic density calculations can be used to optimize dark matter  searches by motivating choices of  parameters where the LHC can probe most deeply into the dark matter parameter space. In the event that DM is seen outside of the region giving the correct relic abundance, we will learn that either thermal relic DM is ruled out in that model, or the DM-quark coupling is suppressed relative to the DM coupling strength to other SM particles. 
}
\begin{document}

\hfill
SISSA 54/2014/FISI

\maketitle

\section{Introduction}

After the recent discovery of a Standard Model (SM)-like Higgs boson \cite{Aad:2012tfa,Chatrchyan:2012ufa}, the focus of  Run II  at the Large Hadron Collider (LHC),  with $\sqrt{s}=(13-14)$ TeV and  a luminosity of about $10^{34}$ cm$^{-2}$ s$^{-1}$,  is to search for new physics at the TeV scale. In particular, 
one of the most commonly publicized goals is to find evidence of a new fundamental particle which might compose the  so-called Dark Matter (DM), a   non-relativistic degree of freedom contributing to about 30\% of the energy in  our universe. 

If DM consists of particles whose mass and interactions with SM particles are
dictated by physics in the electroweak energy range, there are two bonuses. First, 
the  DM particles  might be produced  at the LHC and subsequently escape the detectors.  This is the main reason why a great deal of effort by the ATLAS and CMS collaborations is dedicated to 
 searching for events where large missing energy is the dominant discriminating signature of
 DM \cite{monojetCMS2, 
monojetATLAS2, Abdallah:2014dma}.
Secondly, the DM abundance in our universe is likely to be fixed by the
thermal freeze-out phenomenon: DM particles, initially present in our universe in thermal equilibrium abundance,  annihilate with one another till chemical equilibrium is lost due to the expansion of the universe \cite{Srednicki:1988ce,Scherrer:1985zt,Gondolo:1990dk,Bertone:2004pz}. 
The present-day relic density of these particles is predictable and, in the simple case of $s$-wave self-annihilation of DM in the early universe, it  comes out to be (in units of the critical energy density of the universe)
\be
\Omega_{\rm DM}h^2\simeq \frac{2\times2.4\times 10^{-10}\,{\rm GeV}^{-2}}{\langle\sigma v\rangle_{\rm ann}},
\label{simplerelic}
\ee
where $\langle\sigma v\rangle_{\rm ann}$ is the total thermally-averaged annihilation cross section,
and the factor of 2 in the numerator is made explicit to emphasize the fact that we are assuming a 
non-self-conjugate DM particle.
This abundance must match the one recently measured by the Planck collaboration, $\Omega_{\rm DM}^{\rm obs}h^2=0.1199\pm0.0027$ \cite{Ade:2013zuv}. 

Now, a fundamental question which one should ask is the following: under the optimistic hypothesis that the next 
LHC run gives evidence for  a new weakly interacting particle with a lifetime that exceeds
 about a microsecond,  how confident can we be in claiming  we have finally revealed  the true nature of the DM?  

To answer this question, one  needs to operate within a given framework and identify the parameter space which is compatible with a positive LHC signal and possibly  with the current (and future) bounds (or signals) from direct and indirect searches. Last, but not least, it is imperative to check if the properties of the new particle are compatible with the  observed DM abundance. 

The goal of this paper is to investigate whether a  new stable particle,  within a given set of models, may be assigned the label of thermal relic DM  by comparing the regions of the parameter space where the right abundance is attained with 
 the  exclusion regions  for   the forthcoming Run II at the LHC. The latter are  
 a useful benchmark for evaluating the sensitivity of
the analysis at 14 TeV. However, if DM  is within the reach of the LHC, it is also useful to make the comparison  with the 5$\sigma$ discovery potential regions.  Of course, one can also reverse the logic of this exercise and  identify  the regions of the parameter space of a given model where the DM abundance fits the observed one.  This might be useful to set priorities for  the 
LHC collaborations when comparing the future data with the plethora of models. 

This is not to say that this analysis can exclude the possibility that a new stable particle can be DM. Rather, if the new particle is inconsistent with thermal-relic DM under our assumptions and in a particular model,  then we learn that either: {\bf 1)} the model is not the correct model of DM, or {\bf 2)} one of the assumptions enumerated in Section~\ref{general} do not hold.

This paper is structured as follows. In Section ~\ref{general} we provide some general considerations and state our assumptions, along with a description of the model we consider.
In Section \ref{EFTresults} we compare ATLAS 14TeV sensitivity with the region of parameter space consistent with thermal relic DM. In Section \ref{SIMPresults} we extend this analysis to simplified models.
Finally, we collect our concluding remarks in Section  \ref{conclusion}.

\section{Working assumptions\label{general}}

The goal of this Section is to provide some general considerations about the DM abundance and its link with collider searches and, above all, to list as clearly as possible the set of assumptions we are working with.

\subsection{DM Abundance Considerations}\label{subsec:DM-Ab-Cons}

Consider the general scenario where a DM candidate $\chi$ will eventually be efficiently pair-produced at the LHC. This implies that $\chi$ must interact with first-generation
quarks, therefore one can define the thermally averaged DM annihilation cross section
\footnote{Gluons and other quarks can of course contribute to DM production at the LHC, so the $_*$ subscript defines a reference  channel rather than all possible channels of DM  production at the LHC.} 
\be
 \langle\sigma v\rangle_* \equiv \sigv_{ \chi \bar\chi\rightarrow u \bar u  } + \sigv_{\chi \bar\chi  \rightarrow d \bar d},
\ee
which also sets a reference for DM production at the LHC. 
In the early universe, besides  annihilations into quarks, there can be additional annihilation
channels, so that the total DM annihilation cross section which is relevant for the relic abundance 
is
\be
\label{f}
\langle\sigma v\rangle_{\rm ann}\geq \langle\sigma v\rangle_{*}.
\ee
So, by requiring that the particles $\chi$ and $\bar\chi$ compose the DM abundance, we find
\be
\Omega_{\rm DM}^{\rm obs} h^2 \simeq \frac{2\times2.4\times 10^{-10}\,{\rm GeV}^{-2}}{\langle\sigma v\rangle_{\rm ann}}\leq  \frac{2\times2.4\times 10^{-10}\,{\rm GeV}^{-2}}{\langle\sigma v\rangle_{*}},
\ee
or 
\be
\label{first}
\langle\sigma v\rangle_{*}\lesssim 4.0\times 10^{-9}\,{\rm GeV}^{-2}.
\ee
On the other hand, one can make the reasonable assumption that the dominant DM annihilation channel is to SM fermions and the coupling to the first generation of quarks is not less than the coupling to other SM fermions. This hypothesis follows from the requirement that the would-be DM particles
are efficiently produced in the next Run II. We are the first to admit that this assumption is debatable, but we consider it as a working hypothesis. We will show later how weakening this assumption affects our results. 
In this case, we get 
\be
\label{s}
\langle\sigma v\rangle_{\rm ann}\leq \sum_{\rm quark\,gen.} \langle\sigma v\rangle_{*} + \sum_{\rm lepton\,gen.} \frac{1}{3} \langle\sigma v\rangle_{*} \simeq 4\langle\sigma v\rangle_{*},
\ee
and therefore

\be
\Omega_{\rm DM}^{\rm obs} h^2 \simeq \frac{2\times2.4\times 10^{-10}\,{\rm GeV}^{-2}}{\langle\sigma v\rangle_{\rm ann}}\gtrsim  \frac{2\times6.0\times 10^{-11}\,{\rm GeV}^{-2}}{\langle\sigma v\rangle_{*}},
\ee
or

\be
\label{second}
\langle\sigma v\rangle_{*}\gtrsim 1.0\times 10^{-9}\,{\rm GeV}^{-2}.
\ee
Let us illustrate the relevance of these inequalities with a simple example.
Assume that the interactions between  DM and SM quarks are described within an Effective Field Theory (EFT), where the basic parameters are the DM mass $m_{\rm DM}$ and the UV scale $\Lambda$. Let us also
imagine that the annihilation controlling the thermal abundance takes place in the $s$-wave. One therefore expects
roughly that $\langle\sigma v\rangle_{*}\simeq 10^{-1}m_{\rm DM}^2/\Lambda^4$. We then obtain, from Eqs.~(\ref{first}) and (\ref{second}),
\be
\label{a}
0.7 \left(\frac{m_{\rm DM}}{10^2\,{\rm GeV}}\right)^{1/2}\, {\rm TeV}\lesssim \Lambda\lesssim 1.0 \left(\frac{m_{\rm DM}}{10^2\,{\rm GeV}}\right)^{1/2}\, {\rm  TeV}.
\ee
This value of the UV  scale needs to be compatible with the one needed to explain the positive DM signature at the LHC. For instance,  if $\Lambda$ turns out to be larger than the value of the lower bound, one concludes that the would-be DM particle has to annihilate in other channels which we do not have control of and therefore it would be difficult, if not impossible,  to assign it the label ``dark matter".

Curves corresponding to the correct relic abundance have been used as a benchmark or comparison for EFT constraints since the early usage of EFTs \cite{Goodman:2010ku,Goodman:2010yf}. However,
these relic density constraints on thermal DM are usually considered not to be robust:  for a given set of parameters, the relic density can be smaller if the cross section is enhanced by inclusion of other annihilation channels, such as annihilation to leptons; conversely, the true relic density can be larger if there is a larger dark sector including other types of DM.
However, under a modest set of assumptions, these constraints can become substantially more powerful. Throughout this analysis, we will assume:

\begin{enumerate}

\item the  DM candidate $\chi$ makes up 100\% of the DM of the universe;

\item the DM annihilation rate is related to the observed density today via the standard thermal production mechanism;

\item the dominant annihilation channel is to SM fermions, via one dark mediator;

\item the DM couples to $u,d$ quarks, so that it can be produced at the LHC;

\item the coupling to the first generation of quarks is no less than the coupling to other SM fermions.
\end{enumerate}
In this situation, the relic density constraint gives a range within which the dark sector parameters 
should lie.
It is clear that assumption 5 is by no means a certainty, and so we will show how our results are sensitive to relaxing this assumption. In the event of a signal, this assumption can instead be used to learn about the flavour structure of a thermal relic model that attempts to explain the signal. If the signal falls into the region where DM would be overproduced, then there must be enhanced couplings to other SM particles relative to $u,d$ quarks in order to avoid overproduction, or alternatively, the DM is produced by some mechanism other than thermal production.

Assumption 2 can break down if either the DM was not produced thermally in the early universe, or if some other effect breaks the relationship between the DM density and annihilation rate. For example, unusual cosmologies between freezeout and today can influence the relic density of DM \cite{Gelmini:2006pw}.

To summarize, under our generic assumptions 1-5 the DM production cross section must satisfy the bounds 
\be
1.0\times 10^{-9}\,{\rm GeV}^{-2}\simeq \frac{1}{4}\langle\sigma v\rangle_{\rm ann}\leq \langle\sigma v\rangle_{*}\leq \langle\sigma v\rangle_{\rm ann}\simeq 4.0\times 10^{-9}\,{\rm GeV}^{-2}\,,
\ee
where the value of the annihilation cross section is dictated by ensuring the correct relic abundance.

These tidy inequalities break down when we include the effect of the top quark mass, mediator widths, and a more accurate expression for the relic density later in the text, although the principle behind them remains the same. 
The two limits on the cross section describe two contours in the  parameter space:
if $\sigvmin$  is too large, then DM will be underproduced,  we call this the \emph{\color{blue}underproduction line};  
if $\sigvmin$ is too small, then DM will be overproduced; this is called the \emph{\color{orange}overproduction line}. 
This information is summarised in the table below, where $g_{({\rm DM}, f)}$ generically indicate the mediator couplings to DM and SM fermions, respectively.

\begin{center}
  \begin{tabular}{| c | c | l |}
    \hline
    {\bf \color{orange} Overproduction line} & $\sigv_{\rm ann} \simeq 4 \sigvmin$ & 
    \textbf{EFT:}  Max $\Lambda$, min $\mx$. \\
    && \textbf{Simp. model:} Max $M$, min $g_{({\rm DM},f)}$ and $\mx$. \\ 
\hline
    {\bf \color{blue}  Underproduction line} & $\sigv_{\rm ann} = \sigvmin$ & 
    \textbf{EFT:}  Min $\Lambda$, max $\mx$ \\
    &&\textbf{Simp. model:} Min $M$, max $g_{({\rm DM},f)}$ and $\mx$. \\ 
    \hline
  \end{tabular}
\end{center}

\subsection{Models and cross sections\label{SigmaCalcs}}

To illustrate our point, we focus on a class of simplified models where the DM is a  Dirac particle annihilating to SM fermions in the $s$-channel via a $Z'$-type  mediator. This popular scenario has  seen much attention, including searches by CDF \cite{Aaltonen:2008dn}, ATLAS \cite{Aad:2012hf} and CMS \cite{Chatrchyan:2013qha}.

Working with simplified models is more timely than ever. For some years, the use of effective operators has been popular as a way to place general constraints on the dark sector \cite{Cao:2009uw,Beltran:2010ww,Bai:2010hh,Fan:2010gt,Goodman:2010ku, Cheung:2010ua,Zheng:2010js,Cheung:2011nt,Rajaraman:2011wf,Yu:2011by,Goodman:2010yf,Fox:2011pm,Lowette:2014yta}. 
However, there has always been concern that this approximation breaks down at some mediator mass scale and it is now clear that the effective operator assumption is not a good approximation at LHC energies unless the DM-SM coupling is very large \cite{Goodman:2010yf,Fox:2011pm,Fox:2011fx,Goodman:2011jq,Shoemaker:2011vi,Fox:2012ee,Weiner:2012cb,Busoni:2013lha,Busoni:2014sya,Busoni:2014haa, Buchmueller:2013dya}. On the other end of the spectrum, studies of specific well-motivated models such as supersymmetry \cite{Chung:2003fi} or extra dimensions   \cite{ArkaniHamed:1998rs}   continue to play an important role, but the broad parameter space and specific assumptions required in these models make it difficult to draw general conclusions about the dark sector. 
Hence, simplified models have become the best way to constrain the DM  parameter space \cite{Abdallah:2014dma,Malik:2014ggr,Buchmueller:2014yoa,Alves:2011wf,Harris:2014hga,Buckley:2014fba}. However, this parameter space is still broad, and it is usually unfeasible to constrain the entire space in just one analysis. This necessitates a specific choice for one or more parameters -- for example the coupling-strength and mediator-mass may be constrained for a specific choice of the DM mass. Clearly this is sub-optimal, since we do not want our constraints to be valid only for one arbitrary choice of an unknown parameter. It is important to remember that the search for new neutral particles with electroweak couplings is motivated by the existence of dark matter, and so the requirement that these particles  are a viable thermal relic DM candidate can be a powerful motivator for these arbitrary choices. 

There are many other simplified models to choose from. For example, one could consider a model where dark matter couples to the standard model via $s$-channel exchange of a scalar mediator. In these models, the dark sector usually couples to the standard model via mixing between the new dark mediator and the Higgs. This leads to a Yukawa-type mediator-SM coupling, proportional to the SM fermion mass. This suppresses the production rate via $u$ and $d$ quarks at the LHC relative to top quark loop-induced production via gluon initial states. This suppression also applies to the annihilation rate, especially if annihilation to top quarks is kinematically (or otherwise) unavailable, resulting in very large DM masses and couplings and small mediator masses in order to reach the correct relic density. Hence we do not consider this model here. Another alternative is DM coupling to SM particles via exchange of a scalar mediator in the $t$-channel, as studied in e.g. \cite{Bell:2012rg,Chang:2013oia,An:2013xka,Bai:2013iqa,DiFranzo:2013vra}. The phenomenology is a little different here, for example in the t-channel model the colored mediator can decay into a quark-DM pair \cite{Papucci:2014iwa}. Whilst this is an interesting model, we choose to study a $Z'$-type model as it has the best prospective LHC Run-II constraints with which to compare. 

We consider  the general interaction term in the  Lagrangian  for a vector mediator $Z'$,
\begin{equation}
\mathcal{L} = -\sum_f Z'_\mu  [\bar f \gamma^\mu (g^V_f - g^A_f \gamma_5) f] - Z'_\mu \
[\bar \chi \gamma^\mu(\gx^V - \gx^A \gamma_5)\chi],
\label{lagr}
\end{equation}
where $f$ is a generic SM fermion,   
 the kinetic and gauge terms have been omitted, and
 the sum is over the quark and lepton flavours of choice (see e.g.~Ref.~\cite{Buchmueller:2014yoa}).
 
The LHC searches are only mildly sensitive to the ratios  $g^V_f/g^A_f$ and $\gx^V/\gx^A$, however the distinction is important for relic density calculations, and so we consider a pure vector coupling
 ($g^A_{f, \rm{DM}}=0$). In the EFT limit, we also consider pure axial ($g^V_{f, \rm{DM}}=0$) interactions. In the low-energy  limit, 
the Lagrangian (\ref{lagr}) leads to the effective operators
\bea
{\cal O}_V&=&\frac{1}{\Lambda^2}[\bar\chi\gamma^\mu\chi][\bar f\gamma_\mu f]\,,\qquad\qquad\qquad \textrm{(D5)}\\
{\cal O}_A&=&\frac{1}{\Lambda^2}[\bar\chi\gamma^\mu\gamma^5\chi][\bar f\gamma_\mu \gamma^5 f]\,.
\qquad\qquad\textrm{(D8)}
\eea
The effective operators ${\cal O}_V$ and ${\cal O}_A$
correspond to the usual D5 and D8 operators respectively, defined in Ref.~\cite{Goodman:2010ku}.

The process relevant for relic density calculations is the annihilation of DM particles of mass $\mx$ into SM fermions of mass $m_f$
\be
\chi\bar\chi\to f\bar f.
\label{process}
\ee
In the effective operator limit, the relative cross sections per SM fermion flavour, expanded up to order $v^2$, are
\begin{eqnarray}
\label{sigv-EFT-V}(\sigma v)_{*}^V &\simeq& \frac{N_C \mx^2 }{2 \pi \Lambda^4} \left(\sqrt{1-\frac{m_f^2}{\mx^2}}\left(\frac{m_f^2}{\mx^2}+2\right)+v^2\frac{11m_f^4/\mx^4+2m_f^2/\mx^2-4}{24\sqrt{1-m_f^2/\mx^2}}\right),\\
(\sigma v)_{*}^A &\simeq& \frac{N_C}{2 \pi \Lambda^4} \left(m_f^2\sqrt{1-\frac{m_f^2}{\mx^2}}+v^2\frac{23m_f^4/\mx^2-28m_f^2+8\mx^2}{24\sqrt{1-m_f^2/\mx^2}}\right)\label{sigv-EFT-A}.
\end{eqnarray}
where the colour factor $N_C$ is equal to 3 for quarks and 1 for colourless fermions. 
The full expressions relative to the process (\ref{process}) with $Z'$ exchange, and the corresponding mediator widths, are reported in Appendix~\ref{app:cross-sections}. 

\section{Results: Effective operator limit\label{EFTresults}}
In the extreme EFT  limit, for massless SM annihilation products, the annihilation cross section for a dimension-6 operator goes like $\gx^2 g_f^2 \mx^2 / M^4\equiv  \mx^2 / \Lambda^4$, where  $M$ is the mediator mass, and $g_f$ is its coupling with fermion species $f$. Thus, in general, the underproduction contour is a contour of maximum $\gx$, $g_f$, and $\mx$, and of minimum $M$, and vice-versa for the overproduction contour. Here we compare a range of constraints in the effective operator scenario, where the momentum carried by the mediator is assumed to be small relative to its mass and we define  \footnote{The parameter $\Lambda$ is sometimes called $M_\star$.}
\be
\Lambda \equiv \frac{M}{ \sqrt{\gx g_f}}.
\ee 
The LHC constraints in this scenario are generally valid in the range $\pi \lesssim \sqrt{\gx g_f} \lesssim 4\pi$ \cite{ATL-PHYS-PUB-2014-007}. Since the annihilations relevant to relic density calculations take place when the DM is non-relativistic, the effective operator approximation is valid as long as $M \gg 2\mx$, or $\sqrt{\gx g_f} \gg 2\mx/ \Lambda$, while direct detection constraints are valid across the entire parameter space of interest. 

\begin{figure}[t!]
\centering
\includegraphics[width=0.49\textwidth]{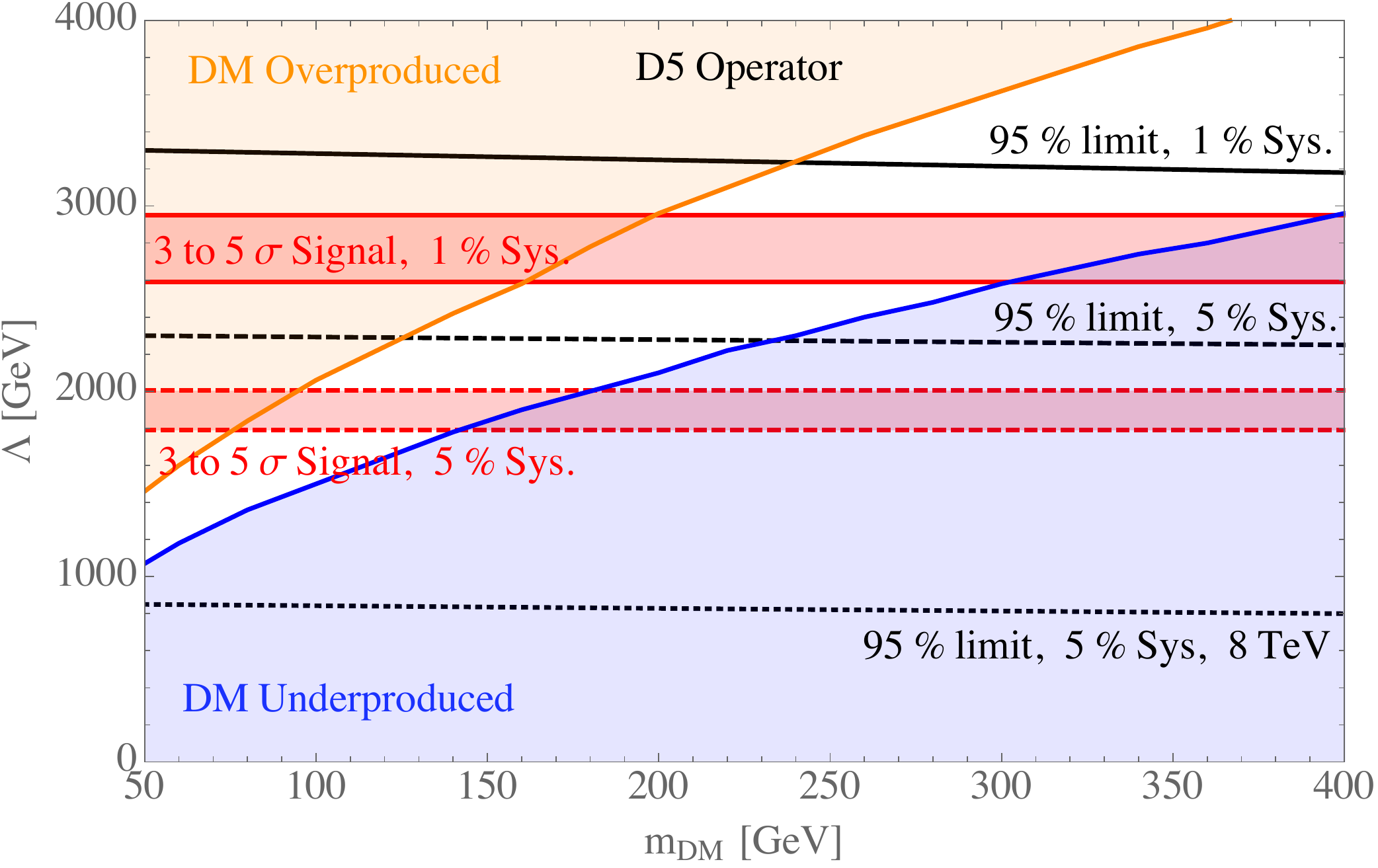}
\includegraphics[width=0.49\textwidth]{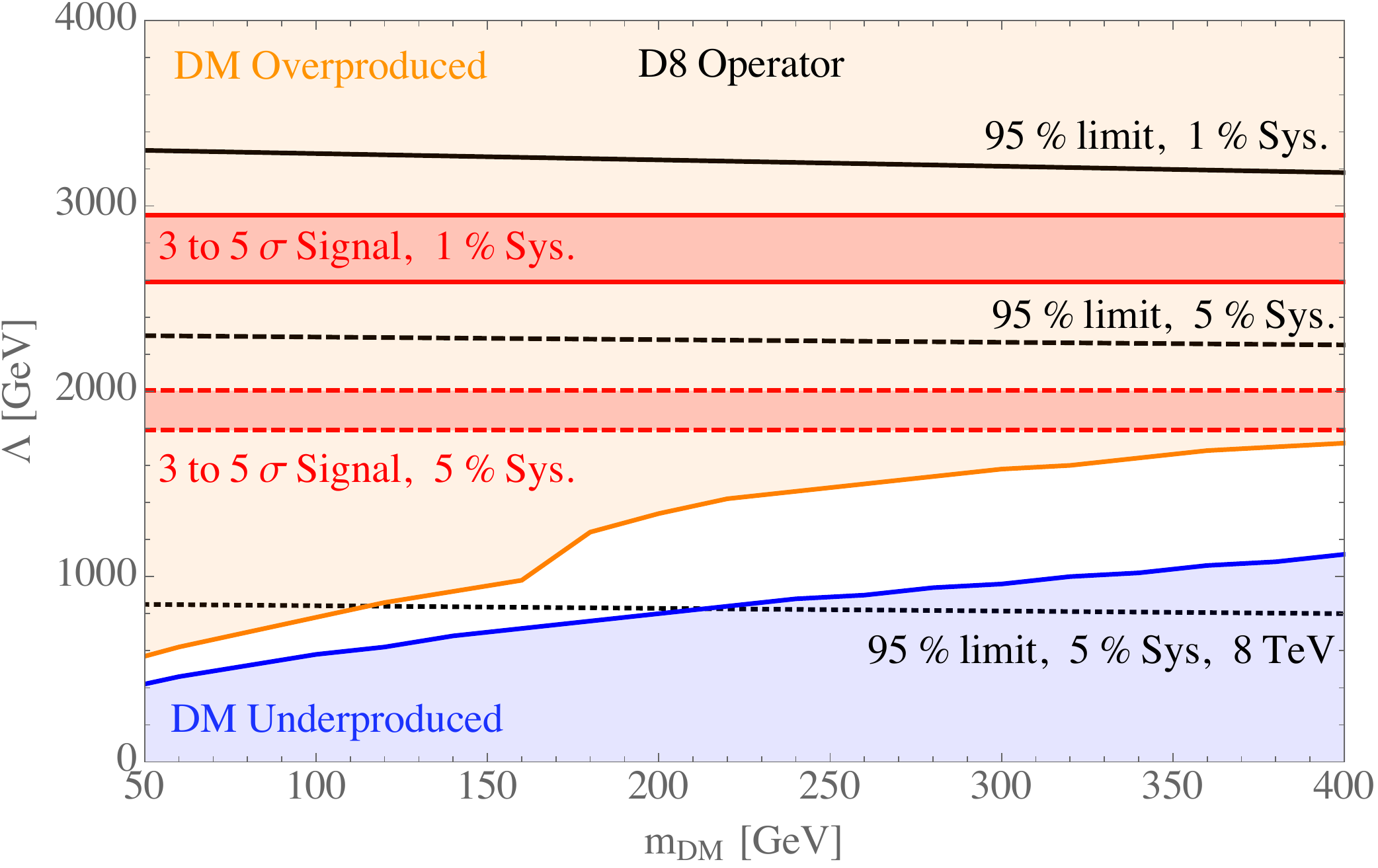}
\caption{
Blue and orange lines show the  under- and over-production lines respectively, defined in the text, for the Vector (D5) (left) and Axial-Vector (D8) (right) operators.  The black lines show prospective ATLAS exclusion limits for various energies and systematic uncertainties, and for luminosities of (3000, 300, 20) fb$^{-1}$ from top to bottom. The red bands show the 3 to 5 $\sigma$ discovery potential \cite{ATL-PHYS-PUB-2014-007}.
EFT approximation is valid for $\pi \lesssim \sqrt{\gx g_f} < 4\pi$ for ATLAS prospects, and $\sqrt{\gx g_f} \gg 2\mx/ \Lambda$ for the relic density constraints. See text for more details. Direct detection constraints are not shown, but for the vector operator D5 they would rule out the entire visible space (cf. sect.~\ref{subsec:dd}).
\label{fig:EFT}}
\end{figure}

Our results in this limit are summarized in Fig.~\ref{fig:EFT}, where we compare the projected exclusion and discovery reach by ATLAS with the \emph{\color{blue}under}- and \emph{\color{orange}over}-production lines defined in the previous Section for the vector and axial-vector operator.
In the following subsections we describe all the elements appearing in Fig.~\ref{fig:EFT}.

\subsection{ATLAS reach}
\label{subsec:ATLreach}

We use simulations of the exclusion and discovery reach of ATLAS at 14 TeV from Ref.~\cite{ATL-PHYS-PUB-2014-007}. This reference estimates the sensitivity of ATLAS to DM in the missing energy + jets channel.  This is a powerful general-purpose channel which has led to strong constraints on DM by both ATLAS and CMS at 7 and 8 TeV \cite{monojetATLAS1,Chatrchyan:2012me}. Searches for other final states such as mono-W/Z \cite{Bell:2012rg,Carpenter:2012rg,ATLASWZ,Aad:2014vka}, mono-photon \cite{monogammaATLAS1,monogammaCMS1}, mono-higgs \cite{Petrov:2013nia,Carpenter:2013xra} and mono-top \cite{Andrea:2011ws} can play a complementary role, especially when combined into mono-all searches. However, the monojet searches still give the strongest constraints \cite{Zhou:2013fla}, and thus represent a good choice for sensitivity studies.

The limits from Ref.~\cite{ATL-PHYS-PUB-2014-007} are only given for two DM masses, $\mx = \{ 50, 400 \}$ GeV, however there is minimal variation in the constraint between the two masses, so we interpolate constraints on $\Lambda$ between these two points.\footnote{We thank Steven Schramm for discussions on this point.} 
These limits are determined for the vector operator, but are expected to be the same for the axial-vector operator \cite{monojetATLAS1}.

The 1\% and 5\% labels indicate projected limits assuming a 1\% or 5\% systematic uncertainty in the SM background, respectively. Achieving 1\% systematics may be overly optimistic, and can be considered a ``best-case scenario''. Other labels indicate the results at a given collision energy and integrated luminosity.
The red bands indicate the potential significance of an observed signal, from 3$\sigma$ to 5$\sigma$. 

\subsection{Direct Detection constraints}
\label{subsec:dd}

We use corrected versions of the  equations from Ref.~\cite{Goodman:2010ku} to translate limits on the spin-dependent (SD) and spin-independent (SI) cross sections into limits on the effective operator parameter $\Lambda$. In this mass range, the  strongest limits are currently from LUX \cite{Akerib:2013tjd} (SI cross section) and Xenon100 \cite{Aprile:2013doa} (SD cross section).  For our simplified models, constraints on $\Lambda$ correspond to a constraint on $M/\sqrt{\gx g_f}$.  

The vector operator ${\cal O}_V$ is subject to constraints on the spin-independent scattering cross section. These constraints are significantly stronger than prospective LHC bounds on this operator, ruling out the entire region displayed in Fig.~\ref{fig:EFT} (left). However, the strength of direct detection constraints falls of quickly below $\mx \simeq 10$ GeV, while LHC constraints are expected to be relatively flat below $\mx = 50$ GeV. If the prospective LHC constraints in Fig.~\ref{fig:EFT} (left) can be extrapolated down, they will become stronger than direct detection constraints at around $\mx = 10$ GeV.
Conversely, the axial-vector  operator ${\cal O}_A$ is subject to much weaker constraints on the spin-dependent scattering cross section. In this range they are barely distinguishable from the $\Lambda=0$ line and thus are not shown.

\subsection{Relic Density Bounds}

In Fig.~\ref{fig:EFT}, we show the \emph{\color{blue}under}- and \emph{\color{orange}over}-production lines defined in the previous Section, for the vector (${\cal O}_V$, D5) and axial-vector (${\cal O}_A$, D8) operator,
under the assumptions 1-5 of Sect.~\ref{subsec:DM-Ab-Cons}.
The range between the orange and blue lines shows the region of parameter space in which any observed $\chi$ can also be thermal relic DM. This marks a good starting point for WIMP searches.
For example, we can see that pure vector DM will be difficult to observe for larger DM masses, and in any case it is ruled out by direct detection constraints. Conversely, axial-vector DM is unconstrained by direct detection,  but it is already heavily constrained by 8 TeV collider bounds, and it is accessible to the 14 TeV searches even for DM masses above 500 GeV.
The jump in the orange line is the point where annihilation into top quarks becomes kinematically allowed.

\begin{figure}[t!]
\centering
\includegraphics[width=0.49\textwidth]{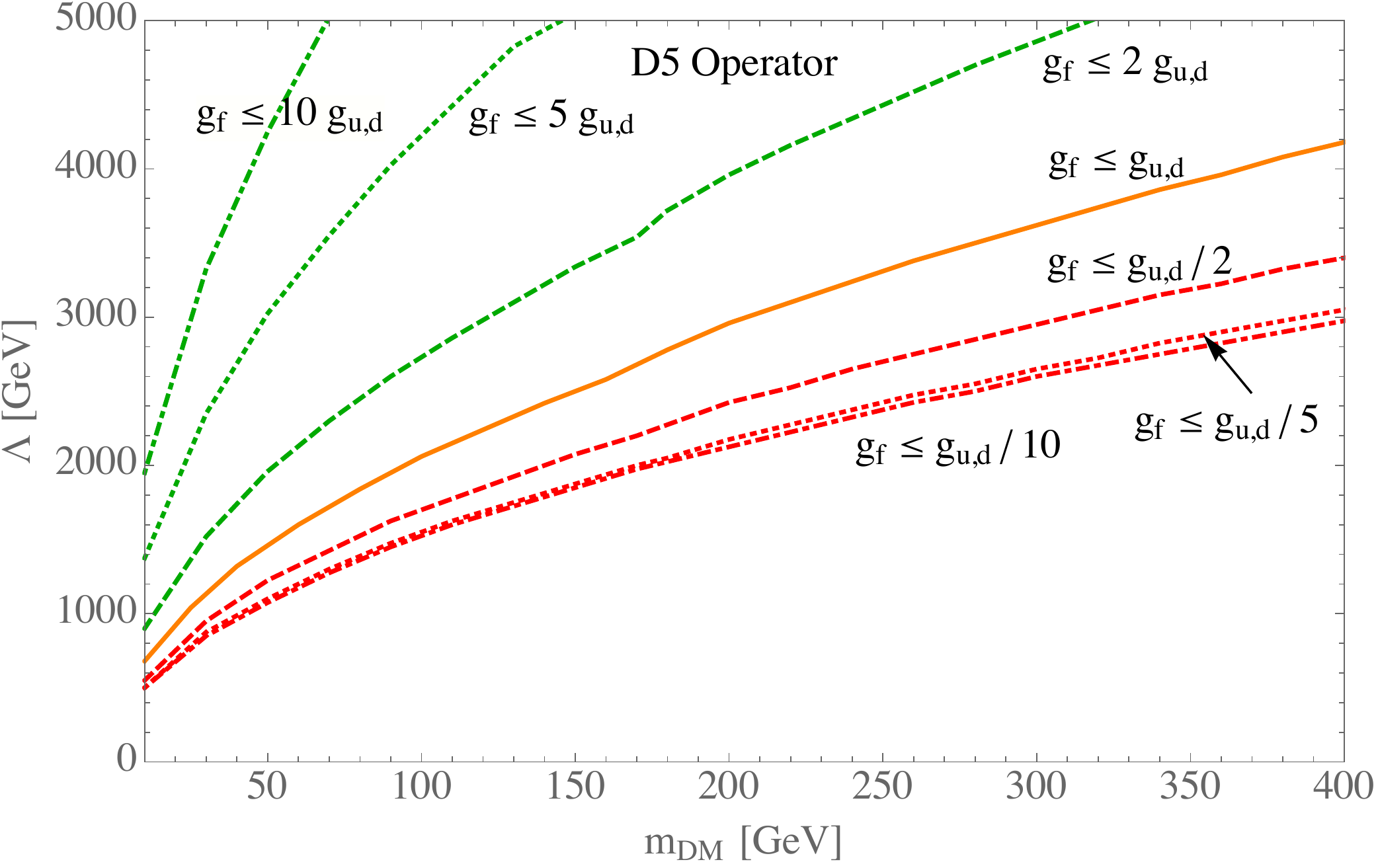}
\includegraphics[width=0.49\textwidth]{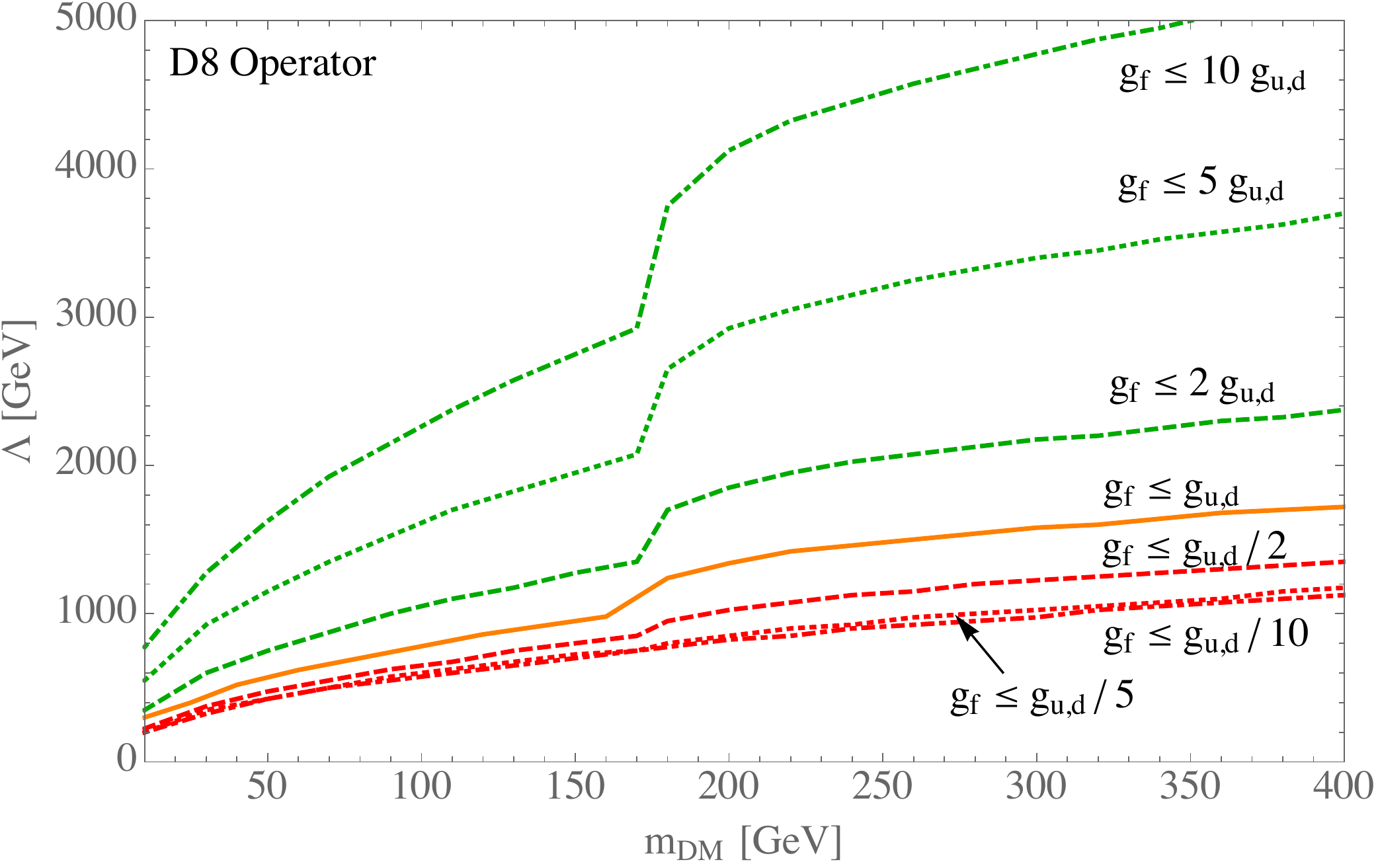}
\caption{
Analogues of the orange overproduction line in Fig. \ref{fig:EFT_scan} (corresponding to $g_f\leq g_{u,d}$), changing the relative value of the coupling between $u,d$ quarks and other SM fermions.
\label{fig:EFT_scan}}
\end{figure}

The overproduction lines in Fig.~\ref{fig:EFT} rely on the assumption  that the DM coupling to the first generation of quarks is not less than the coupling to other SM fermions  ($g_f\leq g_{u,d}$), while the underproduction line only depends on the couplings  $g_{u,d}$ to the first-generation quarks. 
Relaxing/strengthening the assumption 5 of Sect.~\ref{subsec:DM-Ab-Cons} means
allowing the couplings to other SM fermions to span over a  wider/smaller range and correspondingly
 the upper limit is Eq.~(\ref{second}) is changed.
The  effect  on the overproduction lines is shown in Fig.~\ref{fig:EFT_scan}.
We see that if the constraint on $g_f$  is relaxed, the orange line of Fig.~\ref{fig:EFT} gradually becomes too strong, and correspondingly the region in which to search for DM becomes broader
(green curves of Fig.~\ref{fig:EFT_scan}). In the event that a signal compatible with DM is observed, the region where it falls on the plot can be used to infer something about its nature. To be a credible DM candidate, it must either have a production mechanism aside from the usual thermal production, or its couplings to other SM particles can be inferred from where its parameters fall on this plot.

It is also interesting to note that in the EFT limit  there exist simple expressions relating the DM parameters to the correct relic density. In fact, one can trade the annihilation
cross section for the relic abundance (see e.g.~Eq.~(\ref{simplerelic})).
Then, using the Eqs.~(\ref{sigv-EFT-V}) and (\ref{sigv-EFT-A}) for the annihilation cross section in the EFT limit, we can find a simple expression for  a combination of the effective parameters in the 
low-energy theory
\begin{eqnarray}
 \frac{\mx^2}{\Lambda^4} \times\sum_i N^C_i&\simeq& 1\times 10^{-8} \,{\rm GeV}^{-2}\quad (V), \\ 
\frac{m_f^2 + \mx^2 v^2 / 3}{\Lambda^4}\times\sum_i N^C_i &\simeq&  2.5\times 10^{-8} \,{\rm GeV}^{-2}\quad (A),
\end{eqnarray}
where the sum is over the fermionic annihilation products and the colour factor $N^C_i$ is 3 for coloured fermions and 1 for colourless fermions, and this equation assumes $\Lambda$ is the same for all channels.

\section{Results: Simplified models\label{SIMPresults}}

We again consider the two scenarios discussed in the previous section: DM coupling to the minimum and maximum number of SM particles. Now we relax the assumptions leading to the effective operator approximation, and consider the simple UV-complete model described by the Lagrangian (\ref{lagr}). This expands the relevant parameter space from just two parameters, $\mx$ and $\Lambda$, to the set of parameters $\{\mx, M,  \gx^V, \gx^A, g_f^V, g_f^A\}$, where $f$ runs over all SM fermions which the mediator can decay into. As already anticipated, we restrict our attention to the case of pure vector couplings, for which ATLAS projected limits exist \cite{ATL-PHYS-PUB-2014-007}. Thus we consider  $\gx^A  = g_f^A = 0$, and we  define $\gx^V \equiv \gx$, $g_f^V \equiv g_f$. 
The annihilation rates and mediator decay widths have been computed and are shown in Appendix~\ref{app:cross-sections}.

For the \emph{\color{orange} overproduction line}, any change in parameters which decreases the cross section will lead to overproduction of DM.
Similarly, for the \emph{\color{blue} underproduction line},  any change in parameters which increases the cross section will lead to underproduction of DM. 

\begin{figure}[t!]
\centering
\includegraphics[width=0.49\textwidth]{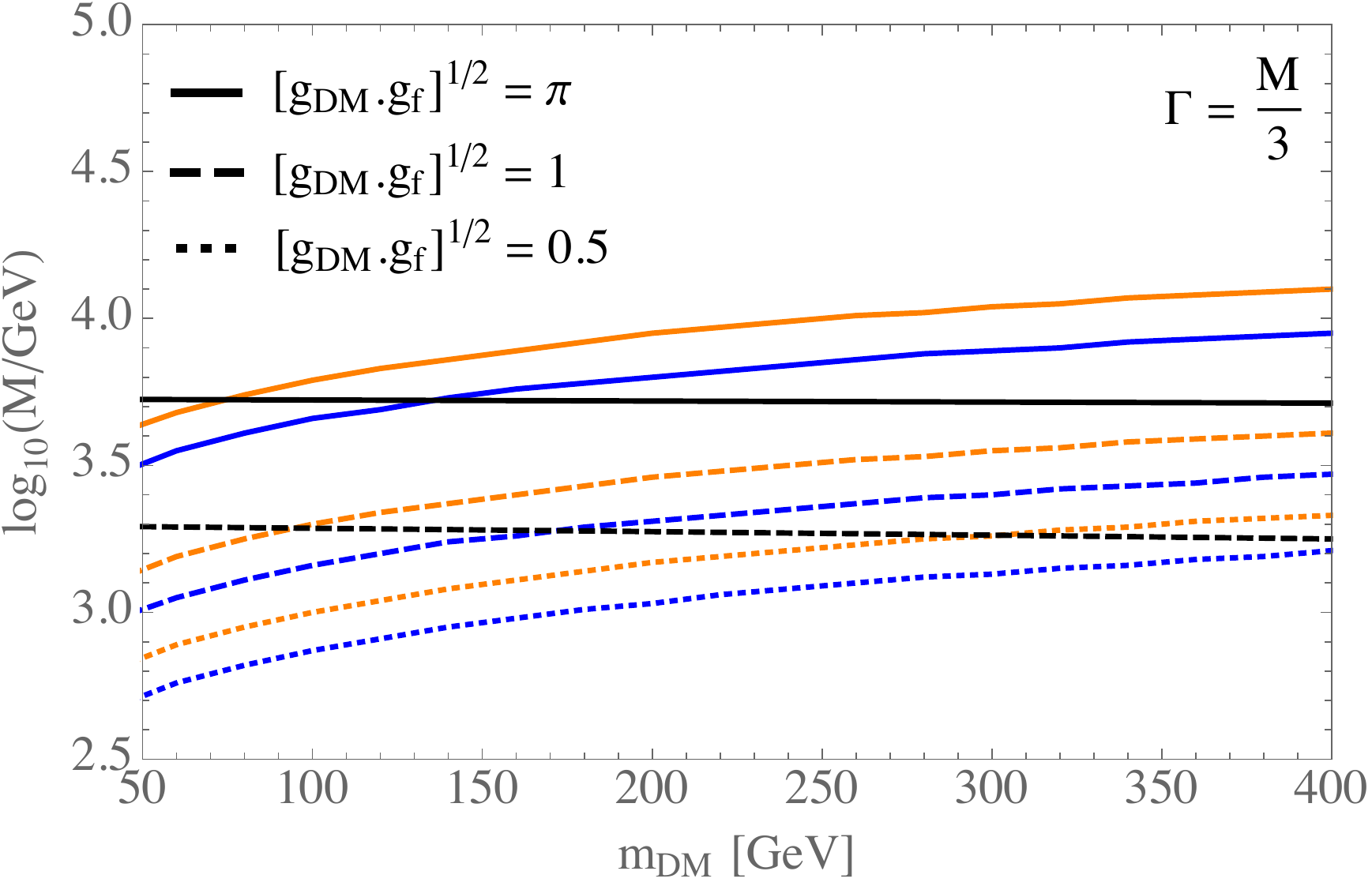}
\includegraphics[width=0.49\textwidth]{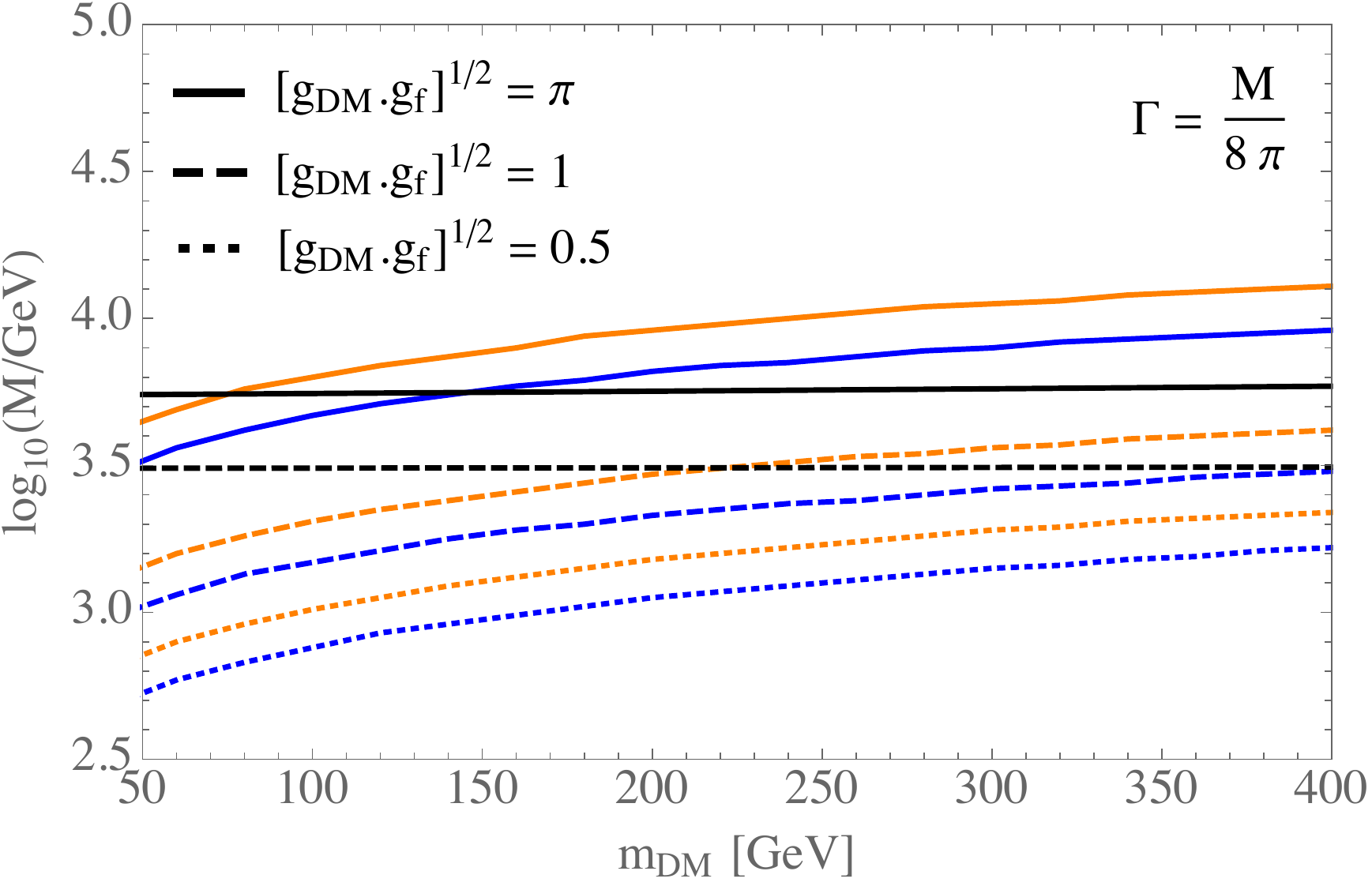}
\caption{
Over- (orange) and under- (blue) production boundary lines for thermal relic dark matter, for three different choices of the coupling strengths, and a $Z'$-type mediator with  pure vector  couplings. Black lines are ATLAS projected 95\% lower bounds after 25 fb$^{-1}$ at 14TeV, assuming 5\% systematic uncertainties. \label{fig:Mvsmx}}
\end{figure}

\begin{figure}[t!]
\centering
\includegraphics[width=0.49\textwidth]{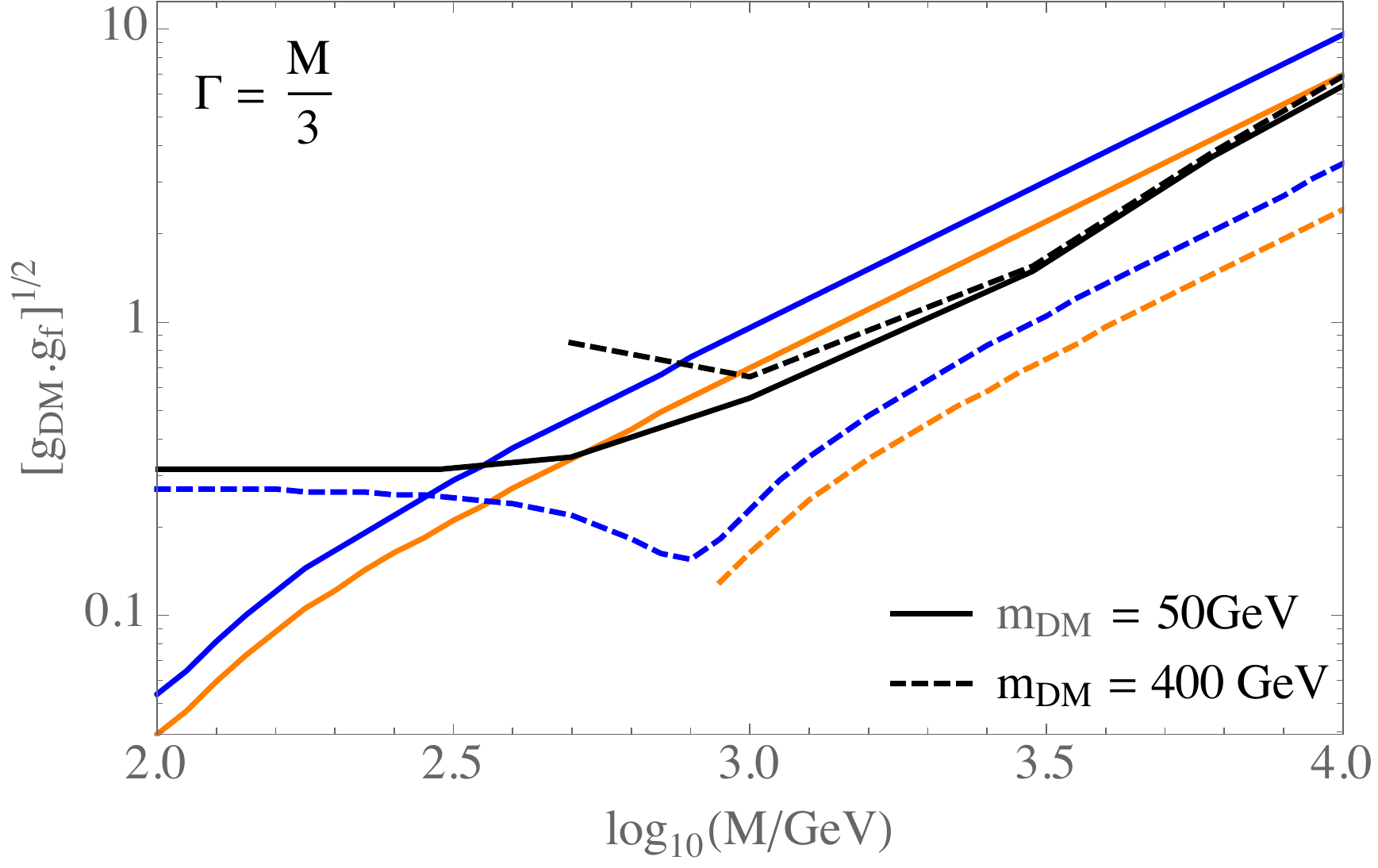}
\includegraphics[width=0.49\textwidth]{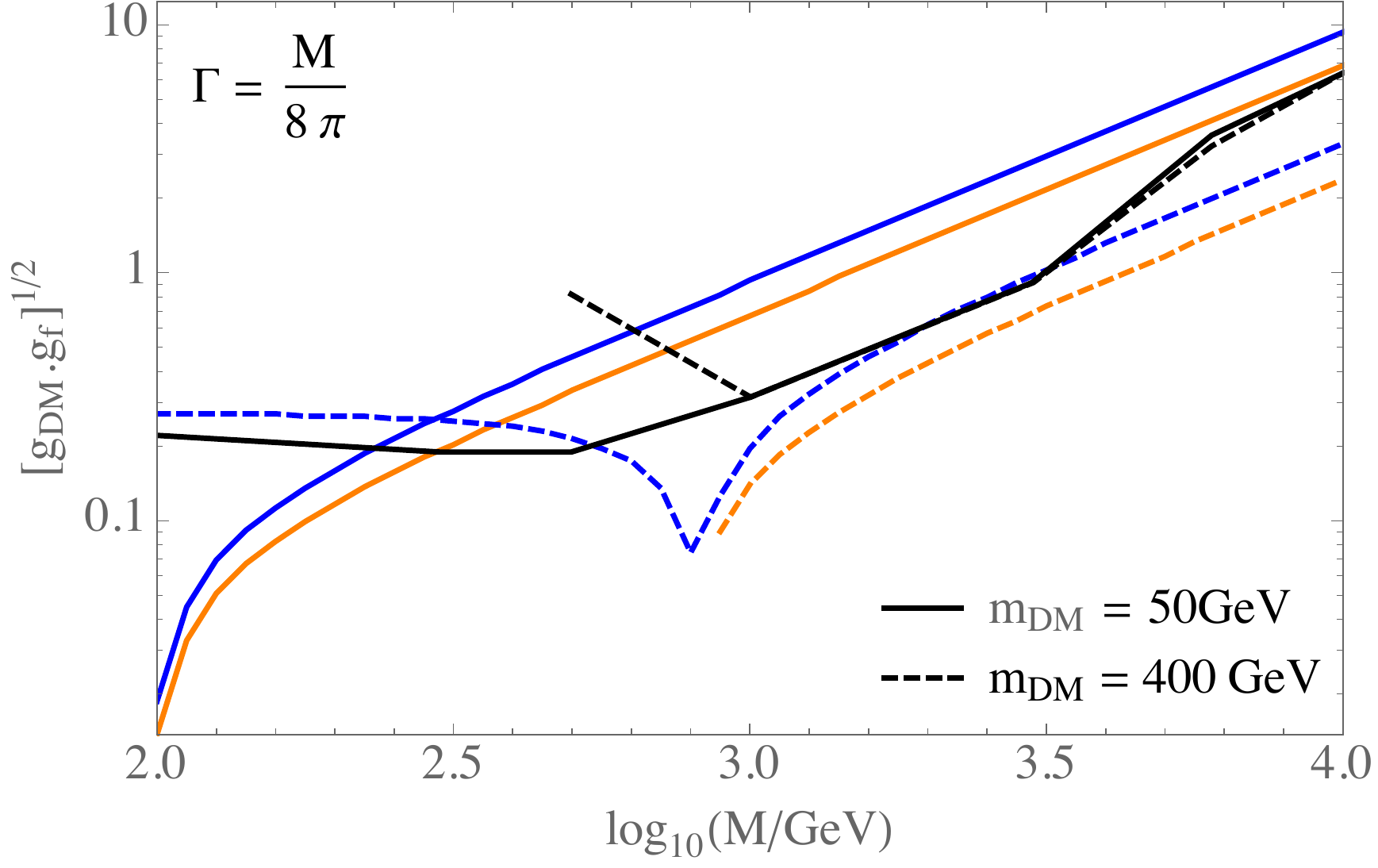}
\caption{
Over- (orange) and under- (blue) production boundary lines for thermal relic dark matter, compared with projected ATLAS reach (black), for two values of the dark matter mass, and a $Z'$-type mediator with  pure vector  couplings. Black lines are ATLAS projected 95\% upper bounds after 25 fb$^{-1}$ at 14TeV, assuming 5\% systematic uncertainties.  
\label{fig:gvsM}}
\end{figure}

In order to compare directly with prospective ATLAS constraints, in Figs.~\ref{fig:Mvsmx}-\ref{fig:gvsM} we show lines for specific choices of $\sqrt{\gx g_f}=$0.5, 1, $\pi$ and $\mx=$50, 400 GeV respectively. The ATLAS constraints are again from Ref.~\cite{ATL-PHYS-PUB-2014-007} and refer to a vector mediator model. These constraints have some degeneracy in $M$ for low values of $\sqrt{\gx g_f}$, and so we do not show a line corresponding to  $\sqrt{\gx g_f}=0.5$
In order to compare with their prospective constraints, the relic density constraints assume the same (arbitrary) widths as ATLAS.

\begin{figure}[t!]
\centering
\includegraphics[width=0.49\textwidth]{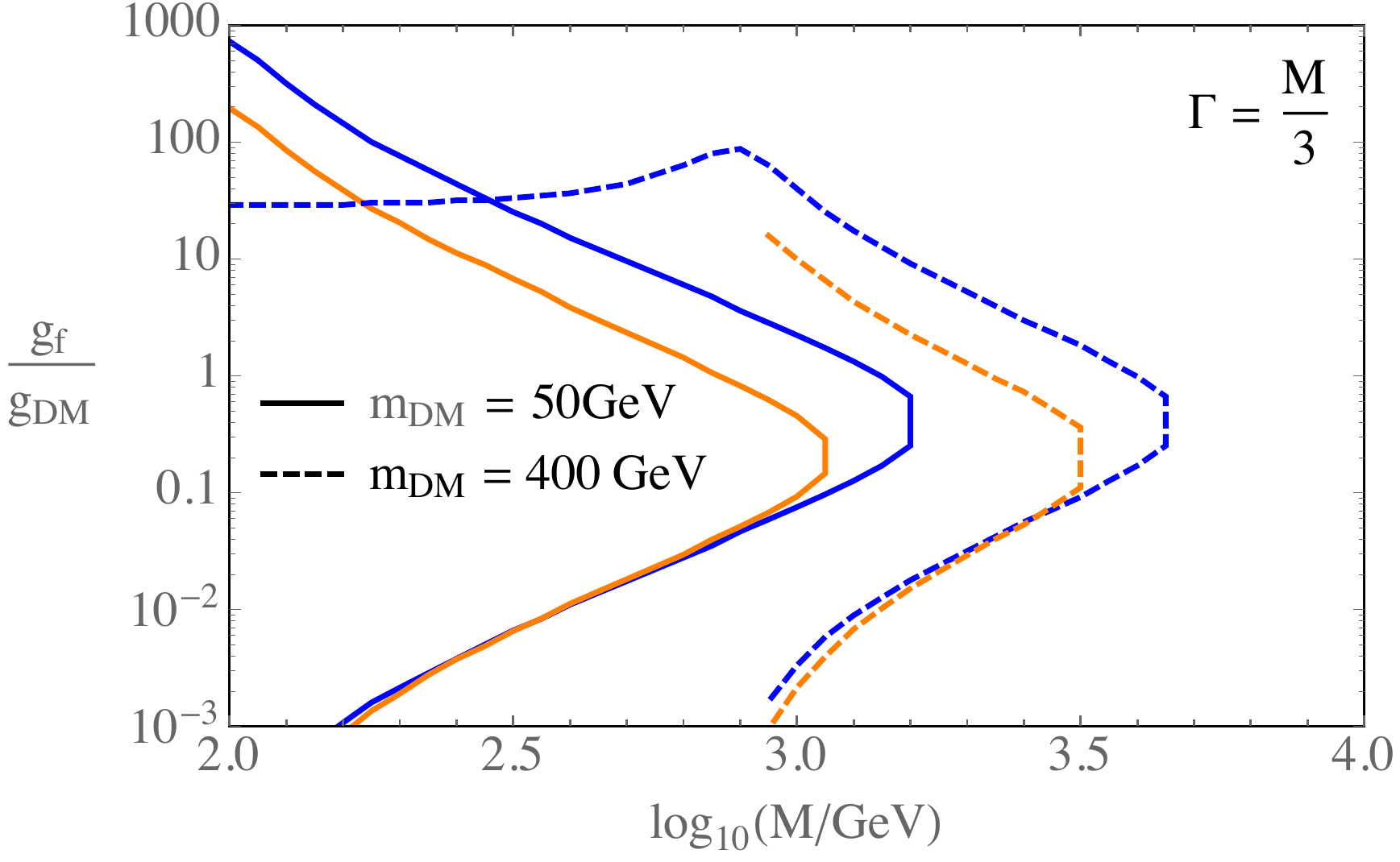}
\includegraphics[width=0.49\textwidth]{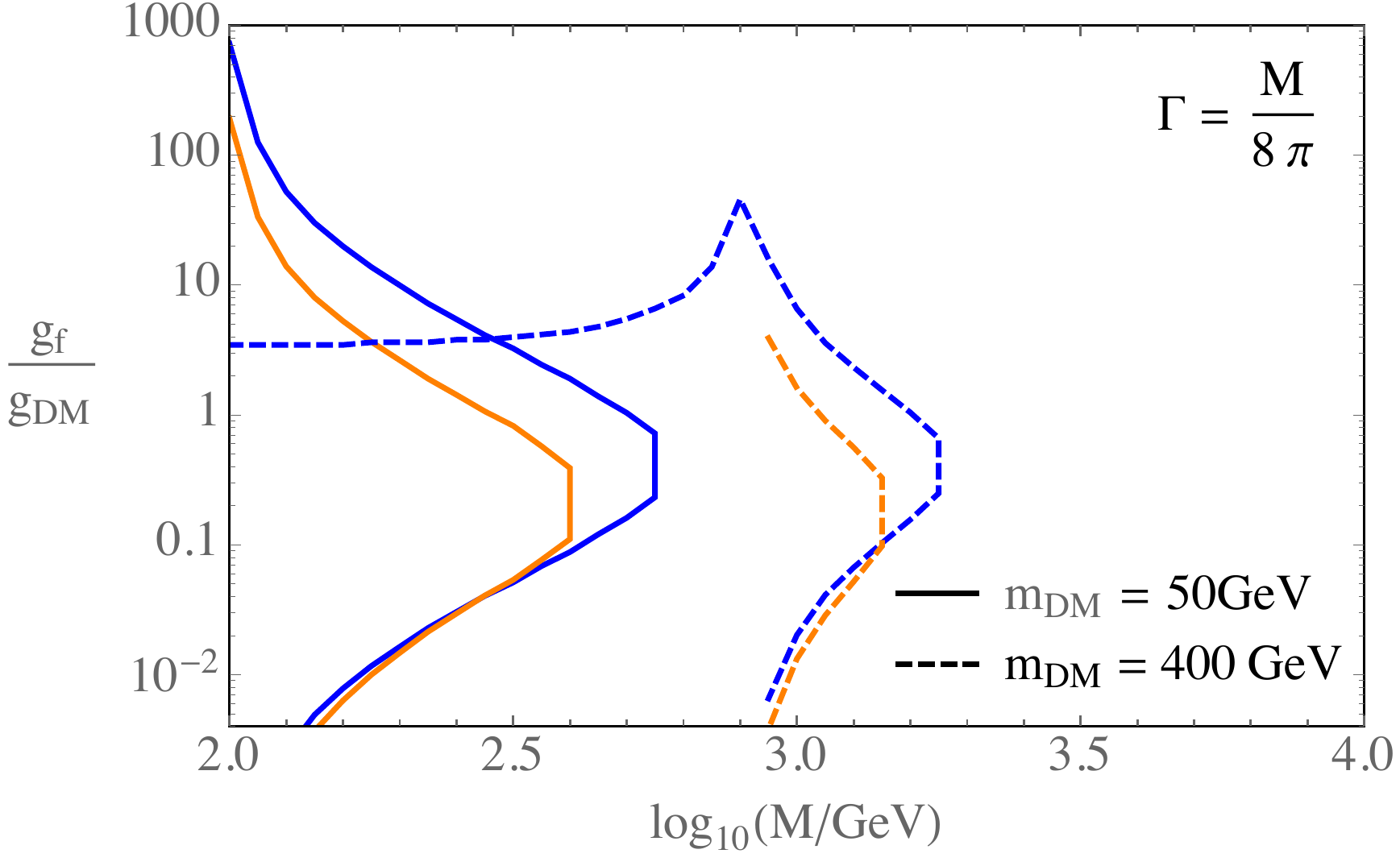}
\caption{
The solution to the ratio $g_f / \gx$ corresponding to the  
bounds on the product $g_f\cdot \gx$ combined with fixed mediator widths (as represented in 
Fig.~\ref{fig:gvsM}). At large mediator masses, no solution exists and the widths are unphysical for the coupling strengths in  Fig.~\ref{fig:gvsM}. \label{fig:ratio}}
\end{figure}

While the annihilation rate of DM particles only depends on the product $\gx g_f$, the mediator decay widths depend on each coupling individually. So we are forced to fix the ratio $g_f / \gx$, in addition to keeping the product $\gx g_f$ as a parameter,.
For fixed values of the mediator width, a bound on the product $\sqrt{\gx g_f}$ can be recast into a bound on the ratio $g_f / \gx$.
The arbitrary widths used in Figs.~\ref{fig:Mvsmx}-\ref{fig:gvsM}  can be compared to the physical widths to fix the ratio  $g_f / \gx$. This is shown in Fig.~\ref{fig:ratio}. In some regions there is no solution, and the width used by ATLAS is in fact not physical.
For this reason we recommend to avoid the use of arbitrary mediator widths, and suggest instead that the widths are fixed to their minimal value given by the decay channels to SM particles and to DM particles.


\section{Conclusion\label{conclusion}}

The upcoming LHC searches for new weakly interacting stable particles may indeed
provide some positive signal in the near future. Then, how confident can we be 
in claiming that the new particle actually accounts for the DM of the universe?

In this paper we have stressed the importance of using relic density considerations
in the searches for DM at the next LHC Run, not only regarded
as a mere constraint but also used as a powerful search tool.
In fact, in order to reveal the true nature of DM,
any future signal of a new weakly interacting
particle possibly produced in the collider must be confronted with the requirement the new particle
has a relic abundance compatible with observations before assigning it the label of thermal DM.

We have followed both the approach of effective operators (in terms of  which most experimental 
analyses are carried out) and the approach of simplified models, for a reference case of a vector mediator. 
We have found that, in both situations, the forthcoming Run II of LHC  has the potential to explore a large 
portion of the parameter space of thermal-relic DM, either in terms of claiming discovery or in terms ruling out models. 

The results of this paper are twofold. One the one hand, they can be used by LHC collaborations as a guidance into the 
parameter space of DM models; in fact,   simple relic density considerations help to set priorities and parameter choices when 
analysing future data in terms of DM.

On the other hand, our results provide clear messages in case of observation of a new stable particle:
if the new particle is not compatible with our thermal relic curves, either it is not the DM or 
one of our working assumptions is not valid. 
In any case, very interesting lessons about the nature of DM will be learned from LHC data.

\section*{Acknowledgments}

We thank A. Boveia, C. Doglioni and S. Schramm for many interesting conversations.
ADS acknowledges partial support from the  European Union FP7  ITN INVISIBLES 
(Marie Curie Actions, PITN-GA-2011-289442).

\appendix

\section{Cross Sections and Widths\label{app:cross-sections}}

In this Appendix we collect the results of cross sections calculations for the process
of DM annihilation into SM fermions
\be
\chi\bar\chi\to f\bar f
\ee
We performed the calculation at zero temperature in the lab frame  where $\chi$  is at rest, and the center of mass energy $s = 2 \mx^2\left(\frac{1}{\sqrt{1-v^2}}+1\right)$. This is equivalent to performing the calculation in the Moeller frame, and is the correct frame for the relic density calculations \cite{Gondolo:1990dk}.

\subsection{Full expressions}

\begin{eqnarray}
(\sigma v)_V &=& \frac{N_C (g_f^V)^2 (\gx^V)^2}{2\pi} \frac{\sqrt{1-m_f^2/\mx^2}}{(M^2 - 4\mx^2)^2 + \Gamma^2 M^2} \Big[(m_f^2+2\mx^2) +
\nonumber\\
&& v^2\Big(\frac{11m_f^4+2m_f^2\mx^2-4\mx^4}{24\mx^2(1-\frac{m_f^2}{\mx^2})}+2\frac{\mx^2(m_f^2+2\mx^2)(M^2-4\mx^2)}{(M^2 - 4\mx^2)^2 + \Gamma^2 M^2}\Big)\Big],\\
(\sigma v)_A &=& \frac{N_C (g_f^A)^2 (\gx^A)^2}{2 \pi}
\frac{\sqrt{1- m_f ^2/ m_{\rm DM} ^2}}{(M^2 - 4 m_{\rm DM} ^2)^2 + \Gamma^2 M^2}
\bigg[\bigg(1 - 4\frac{ m_{\rm DM} ^2}{M^2}\bigg)^2 m_f ^2  +
\nonumber\\
&& v^2\bigg(\Big
(\frac{2m_f^2 m_{\rm DM}^2}{M^4}+\frac{2m_{\rm DM}^2}{M^2}-\frac{3m_f^2}{M^2}+\frac{23m_f^2}{24m_{\rm DM}^2}-\frac{7}{6}\Big)\frac{ m_f ^2}
{(1- m_f ^2/ m_{\rm DM} ^2)}
+\frac{ m_{\rm DM} ^2}
{3(1- m_f ^2/ m_{\rm DM} ^2)}
\nonumber\\
&&+
\frac{2 (M^2 - 4 m_{\rm DM} ^2)^3 m_{\rm DM} ^2 m_f ^2}
{M^4 \left( (M^2 - 4 m_{\rm DM} ^2)^2 + \Gamma^2 M^2 \right)}
\bigg)\bigg].
\end{eqnarray}

\subsection{Limit $m_f \rightarrow 0$}

\begin{eqnarray}
(\sigma v)_V &=& \frac{N_C (g_f^V)^2 (\gx^V)^2}{\pi} \frac{\mx^2}{(M^2 - 4\mx^2)^2 + \Gamma^2 M^2} \left[1 +v^2\left(-\frac{1}{12}+\frac{2\mx^2(M^2-4\mx^2)}{(M^2-4\mx^2)^2+\Gamma^2M^2}\right) \right],\nn\\
&&\\
(\sigma v)_A &=& \frac{N_C (g_f^A)^2 (\gx^A)^2}{6 \pi} \frac{\mx^2}{(M^2 - 4\mx^2)^2 + \Gamma^2 M^2} v^2.
\end{eqnarray}

\subsection{Effective Operator Approximation}

\begin{eqnarray}
(\sigma v)_V &=& \frac{N_C \mx^2 }{2 \pi \Lambda^4} \sqrt{1-\frac{m_f^2}{\mx^2}}\left[\left(\frac{m_f^2}{\mx^2}+2\right)+v^2\frac{11m_f^4/\mx^4+2m_f^2/\mx^2-4}{24(1-m_f^2/\mx^2)}\right],\\
(\sigma v)_A &=& \frac{N_C}{2 \pi \Lambda^4}\sqrt{1-\frac{m_f^2}{\mx^2}} \left[m_f^2+v^2\frac{23m_f^4/\mx^2-28m_f^2+8\mx^2}{24(1-m_f^2/\mx^2)}\right].
\end{eqnarray}

\subsection{Widths}

The widths for the vector mediator decay to a pair of fermions  are given by
\begin{eqnarray}
\Gamma_V & = & \frac{N_C (g_f^V)^2 (M^2 + 2m_f^2) \sqrt{1 - 4m_f^2/M^2}}{12 \pi M}, \\
\Gamma_A & = & \frac{N_C (g_f^A)^2 M (1 - 4m_f^2/M^2)^{3/2}}{12 \pi}.
\end{eqnarray}

\section{Relic density general formalism}

Our technique to compute the abundance and notation follow Refs.~\cite{Gondolo:1990dk} and \cite{Bertone:2004pz}. First we find the freezeout temperature by solving

\begin{equation}
e^{x_F} = \frac{\sqrt{\frac{45}{8}} g_{\rm DoF} \mx M_{\rm Pl} c(c+2) \sigv}{2 \pi^3 g_\star^{1/2} \sqrt{x_F}},
\end{equation}
where $x = \mx / T$ and subscript $F$ denotes the value at freezeout, $g_{\rm DoF} = 2$ is the number of degrees of freedom of the DM particle,
$c$ is a matching constant usually taken to be $1/2$, $g_\star$  is the number of relativistic degrees of freedom, $M_{\rm Pl} = 1/\sqrt{G_N}$ is the Planck mass.
Usually, it is safe to expand in powers of the velocity and use the approximation
\begin{equation}
\sigv = a + b \langle v^2 \rangle + {\cal O}(\langle v^4 \rangle)  \simeq a + 6b/x_F.
\end{equation} 
However, when the mediator width is small, this approximation can down near the $s$-channel resonance in the annihilation rate at $M \simeq 2\mx$ \cite{Griest:1989wd,Griest:1990kh,Gondolo:1990dk} if the width is small. Around this point it becomes more accurate to use the full expression,
\begin{equation}
\sigv = \frac{x}{8 \mx^5 K_2^2[x]}\int_{4\mx^2}^\infty {\rm d}s\, \sigma (s-4\mx^2) \sqrt{s} K_1[\sqrt{s} \,x/\mx].
\end{equation}
With this information, one can calculate the relic abundance,
\begin{equation}
\Omega_{\rm DM}h^2  = \Omega_\chi h^2 + \Omega_{\bar\chi}h^2= \frac{2\times 1.04\times 10^9 \, {\rm GeV}^{-1} \mx}{M_{\rm Pl} \int_{T_0}^{T_F} g_\star^{1/2} \sigv {\rm d}T},\label{relic-full}
\end{equation}
where the factor of 2 is for Dirac DM.
When the non-relativistic approximation to the annihilation rate holds, this simplifies to
\begin{equation}
\Omega_{\rm DM} h^2  = \frac{2\times 1.04\times 10^9 \, {\rm GeV}^{-1} x_F}{\overline{g_\star}^{1/2}  \, M_{\rm Pl} \,(a + 3 b / x_F)}
\label{relic-simple}
\end{equation}
where $\overline{g_\star}^{1/2}$ is a typical value of $g_\star^{1/2}(T)$ in the range $T_0\leq T \leq T_F$.
We have tested the validity of this approximation and find that there is a negligible difference to the full relativistic calculation, since the widths we consider are relatively large. If the physical widths are used, then care should be taken that this approximation still holds when the width becomes small, especially when the annihilation rate has a larger $p$-wave component.

\bibliographystyle{JHEP}
\bibliography{relicLHC}

\providecommand{\href}[2]{#2}\begingroup\raggedright\begin{thebibliography}{10}

\bibitem{Aad:2012tfa}
{\bf ATLAS Collaboration} Collaboration, G.~Aad et~al., {\it {Observation of a
  new particle in the search for the Standard Model Higgs boson with the ATLAS
  detector at the LHC}},  {\em Phys.Lett.} {\bf B716} (2012) 1--29,
  [\href{http://xxx.lanl.gov/abs/1207.7214}{{\tt arXiv:1207.7214}}].

\bibitem{Chatrchyan:2012ufa}
{\bf CMS Collaboration} Collaboration, S.~Chatrchyan et~al., {\it {Observation
  of a new boson at a mass of 125 GeV with the CMS experiment at the LHC}},
  {\em Phys.Lett.} {\bf B716} (2012) 30--61,
  [\href{http://xxx.lanl.gov/abs/1207.7235}{{\tt arXiv:1207.7235}}].

\bibitem{monojetCMS2}
{\bf CMS Collaboration} Collaboration, {\it {Search for new physics in monojet
  events in pp collisions at sqrt(s)= 8 TeV}}, .

\bibitem{monojetATLAS2}
{\bf ATLAS Collaboration} Collaboration, {\it {Search for New Phenomena in
  Monojet plus Missing Transverse Momentum Final States using 10fb-1 of pp
  Collisions at sqrt{s}=8 TeV with the ATLAS detector at the LHC}}, .

\bibitem{Abdallah:2014dma}
J.~Abdallah, A.~Ashkenazi, A.~Boveia, G.~Busoni, A.~De~Simone, et~al., {\it
  {Simplified Models for Dark Matter and Missing Energy Searches at the LHC}},
  \href{http://xxx.lanl.gov/abs/1409.2893}{{\tt arXiv:1409.2893}}.

\bibitem{Srednicki:1988ce}
M.~Srednicki, R.~Watkins, and K.~A. Olive, {\it {Calculations of Relic
  Densities in the Early Universe}},  {\em Nucl.Phys.} {\bf B310} (1988) 693.

\bibitem{Scherrer:1985zt}
R.~J. Scherrer and M.~S. Turner, {\it {On the Relic, Cosmic Abundance of Stable
  Weakly Interacting Massive Particles}},  {\em Phys.Rev.} {\bf D33} (1986)
  1585.

\bibitem{Gondolo:1990dk}
P.~Gondolo and G.~Gelmini, {\it {Cosmic abundances of stable particles:
  Improved analysis}},  {\em Nucl.Phys.} {\bf B360} (1991) 145--179.

\bibitem{Bertone:2004pz}
G.~Bertone, D.~Hooper, and J.~Silk, {\it {Particle dark matter: Evidence,
  candidates and constraints}},  {\em Phys.Rept.} {\bf 405} (2005) 279--390,
  [\href{http://xxx.lanl.gov/abs/hep-ph/0404175}{{\tt hep-ph/0404175}}].

\bibitem{Ade:2013zuv}
{\bf Planck Collaboration} Collaboration, P.~Ade et~al., {\it {Planck 2013
  results. XVI. Cosmological parameters}},  {\em Astron.Astrophys.} (2014)
  [\href{http://xxx.lanl.gov/abs/1303.5076}{{\tt arXiv:1303.5076}}].

\bibitem{Goodman:2010ku}
J.~Goodman, M.~Ibe, A.~Rajaraman, W.~Shepherd, T.~M. Tait, et~al., {\it
  {Constraints on Dark Matter from Colliders}},  {\em Phys.Rev.} {\bf D82}
  (2010) 116010, [\href{http://xxx.lanl.gov/abs/1008.1783}{{\tt
  arXiv:1008.1783}}].

\bibitem{Goodman:2010yf}
J.~Goodman, M.~Ibe, A.~Rajaraman, W.~Shepherd, T.~M. Tait, et~al., {\it
  {Constraints on Light Majorana dark Matter from Colliders}},  {\em
  Phys.Lett.} {\bf B695} (2011) 185--188,
  [\href{http://xxx.lanl.gov/abs/1005.1286}{{\tt arXiv:1005.1286}}].

\bibitem{Gelmini:2006pw}
G.~B. Gelmini and P.~Gondolo, {\it {Neutralino with the right cold dark matter
  abundance in (almost) any supersymmetric model}},  {\em Phys.Rev.} {\bf D74}
  (2006) 023510, [\href{http://xxx.lanl.gov/abs/hep-ph/0602230}{{\tt
  hep-ph/0602230}}].

\bibitem{Aaltonen:2008dn}
{\bf CDF Collaboration} Collaboration, T.~Aaltonen et~al., {\it {Search for new
  particles decaying into dijets in proton-antiproton collisions at s**(1/2) =
  1.96-TeV}},  {\em Phys.Rev.} {\bf D79} (2009) 112002,
  [\href{http://xxx.lanl.gov/abs/0812.4036}{{\tt arXiv:0812.4036}}].

\bibitem{Aad:2012hf}
{\bf ATLAS Collaboration} Collaboration, G.~Aad et~al., {\it {Search for
  high-mass resonances decaying to dilepton final states in pp collisions at
  s**(1/2) = 7-TeV with the ATLAS detector}},  {\em JHEP} {\bf 1211} (2012)
  138, [\href{http://xxx.lanl.gov/abs/1209.2535}{{\tt arXiv:1209.2535}}].

\bibitem{Chatrchyan:2013qha}
{\bf CMS Collaboration} Collaboration, S.~Chatrchyan et~al., {\it {Search for
  narrow resonances using the dijet mass spectrum in pp collisions at
  $\sqrt{s}$=8  TeV}},  {\em Phys.Rev.} {\bf D87} (2013), no.~11 114015,
  [\href{http://xxx.lanl.gov/abs/1302.4794}{{\tt arXiv:1302.4794}}].

\bibitem{Cao:2009uw}
Q.-H. Cao, C.-R. Chen, C.~S. Li, and H.~Zhang, {\it {Effective Dark Matter
  Model: Relic density, CDMS II, Fermi LAT and LHC}},  {\em JHEP} {\bf 1108}
  (2011) 018, [\href{http://xxx.lanl.gov/abs/0912.4511}{{\tt
  arXiv:0912.4511}}].

\bibitem{Beltran:2010ww}
M.~Beltran, D.~Hooper, E.~W. Kolb, Z.~A. Krusberg, and T.~M. Tait, {\it
  {Maverick dark matter at colliders}},  {\em JHEP} {\bf 1009} (2010) 037,
  [\href{http://xxx.lanl.gov/abs/1002.4137}{{\tt arXiv:1002.4137}}].

\bibitem{Bai:2010hh}
Y.~Bai, P.~J. Fox, and R.~Harnik, {\it {The Tevatron at the Frontier of Dark
  Matter Direct Detection}},  {\em JHEP} {\bf 1012} (2010) 048,
  [\href{http://xxx.lanl.gov/abs/1005.3797}{{\tt arXiv:1005.3797}}].

\bibitem{Fan:2010gt}
J.~Fan, M.~Reece, and L.-T. Wang, {\it {Non-relativistic effective theory of
  dark matter direct detection}},  {\em JCAP} {\bf 1011} (2010) 042,
  [\href{http://xxx.lanl.gov/abs/1008.1591}{{\tt arXiv:1008.1591}}].

\bibitem{Cheung:2010ua}
K.~Cheung, P.-Y. Tseng, and T.-C. Yuan, {\it {Cosmic Antiproton Constraints on
  Effective Interactions of the Dark Matter}},  {\em JCAP} {\bf 1101} (2011)
  004, [\href{http://xxx.lanl.gov/abs/1011.2310}{{\tt arXiv:1011.2310}}].

\bibitem{Zheng:2010js}
J.-M. Zheng, Z.-H. Yu, J.-W. Shao, X.-J. Bi, Z.~Li, et~al., {\it {Constraining
  the interaction strength between dark matter and visible matter: I. fermionic
  dark matter}},  {\em Nucl.Phys.} {\bf B854} (2012) 350--374,
  [\href{http://xxx.lanl.gov/abs/1012.2022}{{\tt arXiv:1012.2022}}].

\bibitem{Cheung:2011nt}
K.~Cheung, P.-Y. Tseng, and T.-C. Yuan, {\it {Gamma-ray Constraints on
  Effective Interactions of the Dark Matter}},  {\em JCAP} {\bf 1106} (2011)
  023, [\href{http://xxx.lanl.gov/abs/1104.5329}{{\tt arXiv:1104.5329}}].

\bibitem{Rajaraman:2011wf}
A.~Rajaraman, W.~Shepherd, T.~M. Tait, and A.~M. Wijangco, {\it {LHC Bounds on
  Interactions of Dark Matter}},  {\em Phys.Rev.} {\bf D84} (2011) 095013,
  [\href{http://xxx.lanl.gov/abs/1108.1196}{{\tt arXiv:1108.1196}}].

\bibitem{Yu:2011by}
Z.-H. Yu, J.-M. Zheng, X.-J. Bi, Z.~Li, D.-X. Yao, et~al., {\it {Constraining
  the interaction strength between dark matter and visible matter: II. scalar,
  vector and spin-3/2 dark matter}},  {\em Nucl.Phys.} {\bf B860} (2012)
  115--151, [\href{http://xxx.lanl.gov/abs/1112.6052}{{\tt arXiv:1112.6052}}].

\bibitem{Fox:2011pm}
P.~J. Fox, R.~Harnik, J.~Kopp, and Y.~Tsai, {\it {Missing Energy Signatures of
  Dark Matter at the LHC}},  {\em Phys.Rev.} {\bf D85} (2012) 056011,
  [\href{http://xxx.lanl.gov/abs/1109.4398}{{\tt arXiv:1109.4398}}].

\bibitem{Lowette:2014yta}
S.~Lowette, {\it {Search for Dark Matter at CMS}},
  \href{http://xxx.lanl.gov/abs/1410.3762}{{\tt arXiv:1410.3762}}.

\bibitem{Fox:2011fx}
P.~J. Fox, R.~Harnik, J.~Kopp, and Y.~Tsai, {\it {LEP Shines Light on Dark
  Matter}},  {\em Phys.Rev.} {\bf D84} (2011) 014028,
  [\href{http://xxx.lanl.gov/abs/1103.0240}{{\tt arXiv:1103.0240}}].

\bibitem{Goodman:2011jq}
J.~Goodman and W.~Shepherd, {\it {LHC Bounds on UV-Complete Models of Dark
  Matter}},  \href{http://xxx.lanl.gov/abs/1111.2359}{{\tt arXiv:1111.2359}}.

\bibitem{Shoemaker:2011vi}
I.~M. Shoemaker and L.~Vecchi, {\it {Unitarity and Monojet Bounds on Models for
  DAMA, CoGeNT, and CRESST-II}},  {\em Phys.Rev.} {\bf D86} (2012) 015023,
  [\href{http://xxx.lanl.gov/abs/1112.5457}{{\tt arXiv:1112.5457}}].

\bibitem{Fox:2012ee}
P.~J. Fox, R.~Harnik, R.~Primulando, and C.-T. Yu, {\it {Taking a Razor to Dark
  Matter Parameter Space at the LHC}},  {\em Phys.Rev.} {\bf D86} (2012)
  015010, [\href{http://xxx.lanl.gov/abs/1203.1662}{{\tt arXiv:1203.1662}}].

\bibitem{Weiner:2012cb}
N.~Weiner and I.~Yavin, {\it {How Dark Are Majorana WIMPs? Signals from MiDM
  and Rayleigh Dark Matter}},  {\em Phys.Rev.} {\bf D86} (2012) 075021,
  [\href{http://xxx.lanl.gov/abs/1206.2910}{{\tt arXiv:1206.2910}}].

\bibitem{Busoni:2013lha}
G.~Busoni, A.~De~Simone, E.~Morgante, and A.~Riotto, {\it {On the Validity of
  the Effective Field Theory for Dark Matter Searches at the LHC}},  {\em
  Phys.Lett.} {\bf B728} (2014) 412--421,
  [\href{http://xxx.lanl.gov/abs/1307.2253}{{\tt arXiv:1307.2253}}].

\bibitem{Busoni:2014sya}
G.~Busoni, A.~De~Simone, J.~Gramling, E.~Morgante, and A.~Riotto, {\it {On the
  Validity of the Effective Field Theory for Dark Matter Searches at the LHC,
  Part II: Complete Analysis for the s-channel}},
  \href{http://xxx.lanl.gov/abs/1402.1275}{{\tt arXiv:1402.1275}}.

\bibitem{Busoni:2014haa}
G.~Busoni, A.~De~Simone, T.~Jacques, E.~Morgante, and A.~Riotto, {\it {On the
  Validity of the Effective Field Theory for Dark Matter Searches at the LHC
  Part III: Analysis for the $t$-channel}},
  \href{http://xxx.lanl.gov/abs/1405.3101}{{\tt arXiv:1405.3101}}.

\bibitem{Buchmueller:2013dya}
O.~Buchmueller, M.~J. Dolan, and C.~McCabe, {\it {Beyond Effective Field Theory
  for Dark Matter Searches at the LHC}},  {\em JHEP} {\bf 1401} (2014) 025,
  [\href{http://xxx.lanl.gov/abs/1308.6799}{{\tt arXiv:1308.6799}}].

\bibitem{Chung:2003fi}
D.~Chung, L.~Everett, G.~Kane, S.~King, J.~D. Lykken, et~al., {\it {The Soft
  supersymmetry breaking Lagrangian: Theory and applications}},  {\em
  Phys.Rept.} {\bf 407} (2005) 1--203,
  [\href{http://xxx.lanl.gov/abs/hep-ph/0312378}{{\tt hep-ph/0312378}}].

\bibitem{ArkaniHamed:1998rs}
N.~Arkani-Hamed, S.~Dimopoulos, and G.~Dvali, {\it {The Hierarchy problem and
  new dimensions at a millimeter}},  {\em Phys.Lett.} {\bf B429} (1998)
  263--272, [\href{http://xxx.lanl.gov/abs/hep-ph/9803315}{{\tt
  hep-ph/9803315}}].

\bibitem{Malik:2014ggr}
S.~Malik, C.~McCabe, H.~Araujo, A.~Belyaev, C.~Boehm, et~al., {\it {Interplay
  and Characterization of Dark Matter Searches at Colliders and in Direct
  Detection Experiments}},  \href{http://xxx.lanl.gov/abs/1409.4075}{{\tt
  arXiv:1409.4075}}.

\bibitem{Buchmueller:2014yoa}
O.~Buchmueller, M.~J. Dolan, S.~A. Malik, and C.~McCabe, {\it {Characterising
  dark matter searches at colliders and direct detection experiments: Vector
  mediators}},  \href{http://xxx.lanl.gov/abs/1407.8257}{{\tt
  arXiv:1407.8257}}.

\bibitem{Alves:2011wf}
{\bf LHC New Physics Working Group} Collaboration, D.~Alves et~al., {\it
  {Simplified Models for LHC New Physics Searches}},  {\em J.Phys.} {\bf G39}
  (2012) 105005, [\href{http://xxx.lanl.gov/abs/1105.2838}{{\tt
  arXiv:1105.2838}}].

\bibitem{Harris:2014hga}
P.~Harris, V.~V. Khoze, M.~Spannowsky, and C.~Williams, {\it {Constraining Dark
  Sectors at Colliders: Beyond the Effective Theory Approach}},
  \href{http://xxx.lanl.gov/abs/1411.0535}{{\tt arXiv:1411.0535}}.

\bibitem{Buckley:2014fba}
M.~R. Buckley, D.~Feld, and D.~Goncalves, {\it {Scalar Simplified Models for
  Dark Matter}},  \href{http://xxx.lanl.gov/abs/1410.6497}{{\tt
  arXiv:1410.6497}}.

\bibitem{Bell:2012rg}
N.~F. Bell, J.~B. Dent, A.~J. Galea, T.~D. Jacques, L.~M. Krauss, et~al., {\it
  {Searching for Dark Matter at the LHC with a Mono-Z}},  {\em Phys.Rev.} {\bf
  D86} (2012) 096011, [\href{http://xxx.lanl.gov/abs/1209.0231}{{\tt
  arXiv:1209.0231}}].

\bibitem{Chang:2013oia}
S.~Chang, R.~Edezhath, J.~Hutchinson, and M.~Luty, {\it {Effective WIMPs}},
  {\em Phys.Rev.} {\bf D89} (2014) 015011,
  [\href{http://xxx.lanl.gov/abs/1307.8120}{{\tt arXiv:1307.8120}}].

\bibitem{An:2013xka}
H.~An, L.-T. Wang, and H.~Zhang, {\it {Dark matter with $t$-channel mediator: a
  simple step beyond contact interaction}},
  \href{http://xxx.lanl.gov/abs/1308.0592}{{\tt arXiv:1308.0592}}.

\bibitem{Bai:2013iqa}
Y.~Bai and J.~Berger, {\it {Fermion Portal Dark Matter}},  {\em JHEP} {\bf
  1311} (2013) 171, [\href{http://xxx.lanl.gov/abs/1308.0612}{{\tt
  arXiv:1308.0612}}].

\bibitem{DiFranzo:2013vra}
A.~DiFranzo, K.~I. Nagao, A.~Rajaraman, and T.~M.~P. Tait, {\it {Simplified
  Models for Dark Matter Interacting with Quarks}},  {\em JHEP} {\bf 1311}
  (2013) 014, [\href{http://xxx.lanl.gov/abs/1308.2679}{{\tt
  arXiv:1308.2679}}].

\bibitem{Papucci:2014iwa}
M.~Papucci, A.~Vichi, and K.~M. Zurek, {\it {Monojet versus rest of the world
  I: t-channel Models}},  \href{http://xxx.lanl.gov/abs/1402.2285}{{\tt
  arXiv:1402.2285}}.

\bibitem{ATL-PHYS-PUB-2014-007}
{\it {Sensitivity to WIMP Dark Matter in the Final States Containing Jets and
  Missing Transverse Momentum with the ATLAS Detector at 14 TeV LHC}},  Tech.
  Rep. ATL-PHYS-PUB-2014-007, CERN, Geneva, Jun, 2014.

\bibitem{monojetATLAS1}
{\bf ATLAS Collaboration} Collaboration, G.~Aad et~al., {\it {Search for dark
  matter candidates and large extra dimensions in events with a jet and missing
  transverse momentum with the ATLAS detector}},  {\em JHEP} {\bf 1304} (2013)
  075, [\href{http://xxx.lanl.gov/abs/1210.4491}{{\tt arXiv:1210.4491}}].

\bibitem{Chatrchyan:2012me}
{\bf CMS Collaboration} Collaboration, S.~Chatrchyan et~al., {\it {Search for
  dark matter and large extra dimensions in monojet events in $pp$ collisions
  at $\sqrt{s}=7$ TeV}},  {\em JHEP} {\bf 1209} (2012) 094,
  [\href{http://xxx.lanl.gov/abs/1206.5663}{{\tt arXiv:1206.5663}}].

\bibitem{Carpenter:2012rg}
L.~M. Carpenter, A.~Nelson, C.~Shimmin, T.~M. Tait, and D.~Whiteson, {\it
  {Collider searches for dark matter in events with a Z boson and missing
  energy}},  {\em Phys.Rev.} {\bf D87} (2013), no.~7 074005,
  [\href{http://xxx.lanl.gov/abs/1212.3352}{{\tt arXiv:1212.3352}}].

\bibitem{ATLASWZ}
T.~A. collaboration, {\it {Search for dark matter pair production in events
  with a hadronically decaying $W$ or $Z$ boson and missing transverse momentum
  in $pp$ collision data at $\sqrt{s}=8$ TeV with the ATLAS detector}}, .

\bibitem{Aad:2014vka}
{\bf ATLAS Collaboration} Collaboration, G.~Aad et~al., {\it {Search for dark
  matter in events with a Z boson and missing transverse momentum in pp
  collisions at $\sqrt{s}$=8 TeV with the ATLAS detector}},
  \href{http://xxx.lanl.gov/abs/1404.0051}{{\tt arXiv:1404.0051}}.

\bibitem{monogammaATLAS1}
{\bf ATLAS Collaboration} Collaboration, G.~Aad et~al., {\it {Search for dark
  matter candidates and large extra dimensions in events with a photon and
  missing transverse momentum in $pp$ collision data at $\sqrt{s}=7$ TeV with
  the ATLAS detector}},  {\em Phys.Rev.Lett.} {\bf 110} (2013) 011802,
  [\href{http://xxx.lanl.gov/abs/1209.4625}{{\tt arXiv:1209.4625}}].

\bibitem{monogammaCMS1}
{\bf CMS Collaboration} Collaboration, S.~Chatrchyan et~al., {\it {Search for
  Dark Matter and Large Extra Dimensions in pp Collisions Yielding a Photon and
  Missing Transverse Energy}},  {\em Phys.Rev.Lett.} {\bf 108} (2012) 261803,
  [\href{http://xxx.lanl.gov/abs/1204.0821}{{\tt arXiv:1204.0821}}].

\bibitem{Petrov:2013nia}
A.~A. Petrov and W.~Shepherd, {\it {Searching for dark matter at LHC with
  Mono-Higgs production}},  {\em Phys.Lett.} {\bf B730} (2014) 178--183,
  [\href{http://xxx.lanl.gov/abs/1311.1511}{{\tt arXiv:1311.1511}}].

\bibitem{Carpenter:2013xra}
L.~Carpenter, A.~DiFranzo, M.~Mulhearn, C.~Shimmin, S.~Tulin, et~al., {\it
  {Mono-Higgs: a new collider probe of dark matter}},  {\em Phys.Rev.} {\bf
  D89} (2014) 075017, [\href{http://xxx.lanl.gov/abs/1312.2592}{{\tt
  arXiv:1312.2592}}].

\bibitem{Andrea:2011ws}
J.~Andrea, B.~Fuks, and F.~Maltoni, {\it {Monotops at the LHC}},  {\em
  Phys.Rev.} {\bf D84} (2011) 074025,
  [\href{http://xxx.lanl.gov/abs/1106.6199}{{\tt arXiv:1106.6199}}].

\bibitem{Zhou:2013fla}
N.~Zhou, D.~Berge, and D.~Whiteson, {\it {Mono-everything: combined limits on
  dark matter production at colliders from multiple final states}},  {\em
  Phys.Rev.} {\bf D87} (2013), no.~9 095013,
  [\href{http://xxx.lanl.gov/abs/1302.3619}{{\tt arXiv:1302.3619}}].

\bibitem{Akerib:2013tjd}
{\bf LUX Collaboration} Collaboration, D.~Akerib et~al., {\it {First results
  from the LUX dark matter experiment at the Sanford Underground Research
  Facility}},  \href{http://xxx.lanl.gov/abs/1310.8214}{{\tt arXiv:1310.8214}}.

\bibitem{Aprile:2013doa}
{\bf XENON100 Collaboration} Collaboration, E.~Aprile et~al., {\it {Limits on
  spin-dependent WIMP-nucleon cross sections from 225 live days of XENON100
  data}},  {\em Phys.Rev.Lett.} {\bf 111} (2013), no.~2 021301,
  [\href{http://xxx.lanl.gov/abs/1301.6620}{{\tt arXiv:1301.6620}}].

\bibitem{Griest:1989wd}
K.~Griest and M.~Kamionkowski, {\it {Unitarity Limits on the Mass and Radius of
  Dark Matter Particles}},  {\em Phys.Rev.Lett.} {\bf 64} (1990) 615.

\bibitem{Griest:1990kh}
K.~Griest and D.~Seckel, {\it {Three exceptions in the calculation of relic
  abundances}},  {\em Phys.Rev.} {\bf D43} (1991) 3191--3203.

\end{thebibliography}\endgroup

\end{document}